\documentclass[11pt]{article}
\usepackage[utf8]{inputenc}
\usepackage{amssymb}
\usepackage{amsfonts}
\usepackage{graphicx}
\usepackage{graphics}
\usepackage{color}
\usepackage{fixltx2e}
\usepackage{amsmath}
\usepackage{authblk}
\usepackage{cite}

\title{The role of late photons in diffuse optical imaging}

\author[1]{Jack Radford\thanks{j.radford.1@research.gla.ac.uk}}
\author[1]{Ashley Lyons}
\author[2]{Francesco Tonolini}
\author[1]{Daniele Faccio\thanks{daniele.faccio@glasgow.ac.uk}}
\affil[1]{\textit{School of Physics and Astronomy, University of Glasgow, Glasgow, G12 8QQ, UK}}
\affil[2]{\textit{School of Computing Science, University of Glasgow, Glasgow, G12 8QQ, UK}}

\date{}

\begin{document}
\maketitle

\begin{abstract}
The ability to image through turbid media such as organic tissues, is a highly attractive prospect for biological and medical imaging. This is challenging however, due to the highly scattering properties of tissues which scramble the image information. The earliest photons that arrive at the detector are often associated with ballistic transmission, whilst the later photons are associated with complex paths due to multiple independent scattering events and are therefore typically considered to be detrimental to the final image formation process. In this work we report on the importance of these highly diffuse, ``late'' photons for computational time-of-flight diffuse optical imaging. In thick scattering materials, $>80$ transport mean free paths, we provide evidence that including late photons in the inverse retrieval enhances the image reconstruction quality. We also show that the  late photons alone  have sufficient information to retrieve images of a similar quality to early photon gated data. This result emphasises the importance in the strongly diffusive regime discussed here, of fully time-resolved imaging techniques.
\end{abstract}

\section{Introduction}

Imaging through a highly diffuse medium such as biological tissue, can be severely hindered by the presence of scattering which causes photons to propagate along a random and complicated trajectory. This results in a distortion of the wavefront, where the spatial information is blurred to the extent that an image of a hidden object placed behind, or directly embedded within the scattering medium, cannot be reconstructed from a simple intensity measurement of the transmitted photons. The problem is complicated further since the scattering process causes strong attenuation of transmitted light and hence single photon detection methods may also be required.

Photons travelling in scattering media can be divided into three broad categories \cite{singer1990, wang1991ballistic, Dunsby2003, Yoo:90}: \emph{ballistic photons}, where there is no interaction with the medium and the photons propagate straight through with a coherent wavefront; \emph{snake photons} which undergo a few scattering events; and \emph{diffuse photons} which have scattered many times and have a trajectory that no longer corresponds to their initial propagation direction, leading to an incoherent wavefront. Ballistic and snake photons find many uses for imaging through scattering media \cite{wang1991ballistic,Gopal1999,behind1,Kanaev2018, Tanner:17, Brezner2015}. However, ballistic photons are exponentially suppressed as the propagation depth increases. The average distance a photon can propagate before the trajectory is randomised, i.e. all memory is lost of the initial propagation direction, is characterised by the transport mean free path, $\ell^*$ (TMFP). Media with a propagation length $L \gg \ell^*$ are said to be in the highly diffusive regime and the transmitted light is constituted mostly of diffuse photons. This is the case for many biological tissue media with $\ell^*$ of the order of 0.1 cm and the thickness is of the order of centimetres \cite{tissue-review}.

When imaging in transmission geometry, whereby the light source and detector are on opposite sides of the target, spatial properties of the transmitted photons alone can provide some information about about objects embedded inside a diffusive medium \cite{singer1990}. However, in order to achieve better spatial resolution, effort has been focused on using the temporal profile of the photon arrival statistics. This has led to developments in diffuse optical techniques at the computational and hardware level, including  nonlinear-optical gating approaches \cite{wang1991ballistic, behind1} to more recent advances in single photon detection and reconstruction algorithms \cite{altmann, Chen2010, diffuse-review, Alayed2017}. For weakly scattering media, the simplest approach is to isolate the snake and ballistic photons by time-gating the first photons to arrive at the detector. However, in the highly diffusive regime the strong attenuation of these photons makes this unfeasible. Despite this, the remaining early photons have been used for image reconstruction through a variety of media and experimental configurations \cite{Chen2000,Turner2005, Turner2007}. Fluorescent probes have also been exploited to further enhance the recovered spatial resolution and enable in-vivo measurements \cite{Leblond2009,Ntziachristos2000,Niedre2008, Zhang2011, Pichette2013}. In these approaches, a large portion of the  detected light corresponding to the late-arrival photons, typically chosen as the photons that arrive after the main temporal peak, remain unused for the image reconstruction. 

On the contrary, other time-domain diffuse optical imaging systems, configured in a reflection geometry, use late arriving photons to selectively isolate information from different depths \cite{Torricelli2005, DallaMora2015, DiSieno2016, Lange2018, Selb2006}. However, this does not overcome the limitation of detecting only highly scattered photons when gating for very deep regions of the sample. As such, these techniques are similar to early photon imaging which rely on the, exponentially suppressed, quasi-ballistic photons from targeted depth. The information outwith the time-gate corresponding to photons with a deep explored volume remain unused for spatial reconstructions.

Most work to date focuses on the regime in which the total medium thickness is of order $1-10\ell^*$. Moving beyond this to $30-70\ell^*$, a number of pioneering studies have acknowledged the importance of processing the full photon time-of-arrival information in one form or another in order to improve the retrieved image spatial resolution \cite{delpy95, Hebden1999, Hebden:94, Gao2002, Gandjbakhche1998, Cubeddu1996, Cai1996, Model1997, Zint2003}.

All Photons Imaging (API) has been presented as a technique that also uses the full temporal profile of the transmitted photons to reconstruct images through tissue phantoms \cite{satat2016all}. The technique uses a regularised least-squares optimisation method to match the measured spatial and temporal profile to a predictive forward model. API has shown to yield a significant improvement over the use of time-gated ballistic photons through 1.5 cm of a tissue phantom with the target object positioned on the far side. However, in this demonstration the diffusion was still relatively weak as indeed the  time-integrated image, as well as that retrieved from the ballistic photons, continue to show a strong resemblance to the overall shape of the target object.

The concept of using the full temporal extent of the transmitted light has recently been taken one step further to produce images of 2D objects embedded within 5 cm of tissue phantom \cite{Lyons2019b}. The transmitted photons were detected with a Single Photon Avalanche Diode (SPAD) array with the time-of-flight information provided using Time Correlated Single Photon Counting (TCSPC) at each individual pixel, thus recording both the spatial and temporal properties of the transmitted light. This method also uses regularised least-squares optimisation applied to the full 3-dimensional data for image reconstruction and has been demonstrated to resolve feature sizes on the order of 1 mm through more than 80 transport mean free paths

\begin{figure}[h!]
\centering\includegraphics[width=0.7\linewidth]{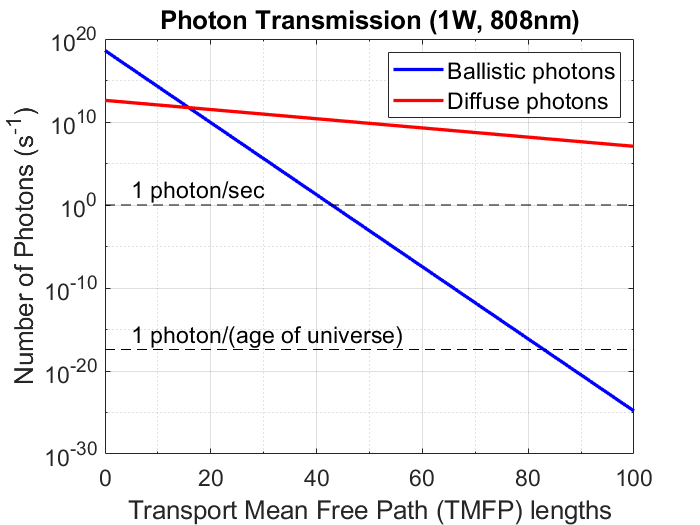}
\caption{The number of transmitted ballistic (blue, based on Eq.~\eqref{Eq:ballistic}) and diffuse (red, based on Eq.~\eqref{Eq:diffuse}) photons per second as the transport mean free path length increases. Early photon imaging techniques are typically used in systems around $10\ell^*$ whereas the  system investigated here is $>80\ell^*$.}
\label{ball-Photons}
\end{figure}

The impact of moving towards the goal of imaging through $100\ell^*$ can be appreciated by assessing the impact of this on ballistic and snake photons. In the diffusive regime, the number of ballistic photons decreases exponentially as described by the Beer-Lambert law \cite{Farsiu2007,Dunsby2003, Sordillo2017, Yaroshevsky2011}:
\begin{equation}\label{Eq:ballistic}
    I = I_0\exp{\left[-\frac{L}{\ell^*}\right]}.
\end{equation}
For a 1 W average input power from the light source, $10^{14}$ 1/s ballistic photons are transmitted through $10\ell^*$, corresponding roughly to readily detectable $\mu$W power levels. At $80\ell^*$, this reduces to  $\sim10^{-17}$ 1/s, corresponding roughly to 1 photon transmitted in the age of the Universe. This is further detailed in Fig.~\ref{ball-Photons} where we plot the number of transmitted ballistic (blue curve) and diffuse photons (red curve) with increasing TMFPs of medium. These estimates clearly indicate that in the $80-100\ell^*$ regime, ballistic photons effectively cease to exist and leads us to question the overall role of early and late photons in transmitting image information. 

Here we investigate the importance of the full time-of-flight distribution of the photons transmitted through $\sim80\ell^*$ of diffuse medium with an embedded object. In particular, we assess the role of photons with later arrival times in the quality of the image reconstruction for computational diffuse optical imaging. Using the experimental data acquired in \cite{Lyons2019b}, we demonstrate that the whole temporal profile of the transmitted photons must be sampled to give the best result and highlight the importance of measuring late photons.

\section{Photon Diffusion}

We consider the regime on which the TMFP, $\ell^*$, is significantly less than the total propagation length within the medium, $L$. We will also from hereon refer to the situation and data described in Ref.~\cite{Lyons2019b} where $\ell^* = 600$ $\mu$m and $L = 5$ cm, corresponding to a total propagation  83 TMFPs. We can thus model the light transport as photon diffusion \cite{yoo1990}. 

The photon flux, $\Phi$ within the medium is described by the photon diffusion equation \cite{Patterson1989,wang2007diffuse,Dehghani2009, Konecky2008, Yamada2014, Durduran2010}:
\begin{equation}\label{Eq:PDE}
     c^{-1}\frac{\partial \Phi(\vec{r}, t)}{\partial t}= S(\vec{r}, t)+ D \nabla^2 \Phi(\vec{r}, t) - \mu_a \Phi(\vec{r}, t),
\end{equation}
where $\vec{r}$ is the radial distance from the source, $t$ is time, $c$ is the speed of light in the medium and $S$ is the power density of the light source. The TMFP can be explicitly described as $\ell^*={1}/{(\mu_s'+\mu_a)}$, where $\mu_a$ and $\mu_{s}'$ are the absorption and reduced scattering coefficients, respectively. The transport mean free path is also implicitly used in the diffusion coefficient,  $D={\ell^*}/{3}={1}/{3(\mu_{s}'+\mu_a)}$. 

The forward model of photon diffusion can be numerically simulated by taking the analytical solution to Eq.~(\ref{Eq:PDE}) \cite{Patterson1989}:
\begin{equation}\label{Eq:PDE_phi}
    \Phi(\vec{r}, t) = \frac{c}{(4\pi Dct)^{3/2}} \exp{\left(-\frac{|\vec{r}|^2}{4Dct}-\mu_act\right)}.
\end{equation}

In the regime where a short laser pulse is incident on a scattering material, $\mu_s'\gg\mu_a$ and $L\gg \ell^*$, the intensity of the diffuse photons transmitted through a scattering medium can be modelled as \cite{Sordillo2017, Yaroshevsky2011}:
\begin{equation}\label{Eq:diffuse}
    I = I_0\frac{\delta\Omega}{4\pi} \exp{\left[-\sqrt{3\mu_a(\mu_s'+\mu_a)}L\right]},
\end{equation}
where ${\delta\Omega}/{4\pi}$ is of the order $10^{-6}$ for our set up, accounts for the solid-angle fraction of light which reaches the lens of the detector.

\section{Experimental Layout}

\begin{figure}[h!]
\centering\includegraphics[width=\linewidth]{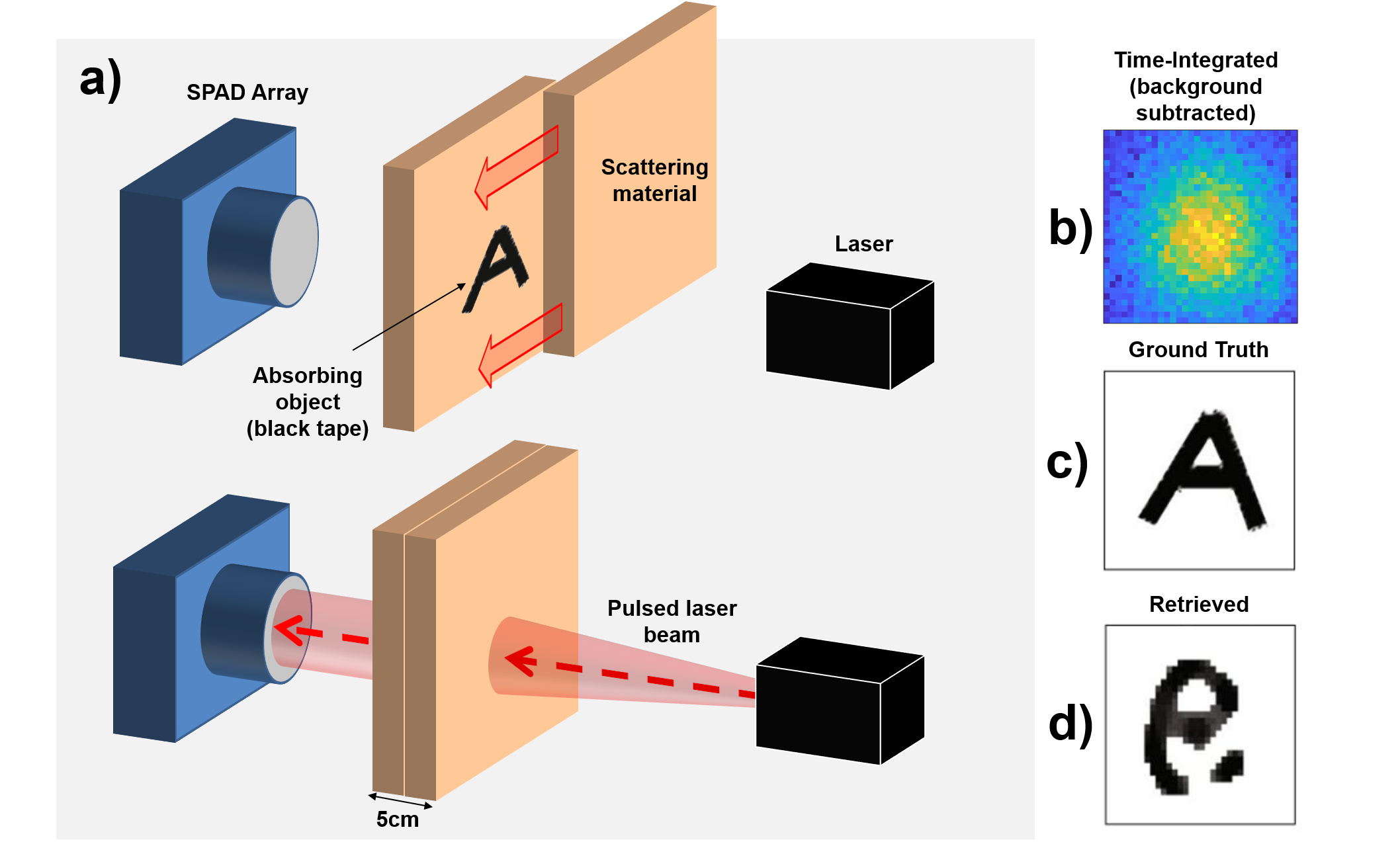}
\caption{a) Layout of the experiment. A hidden 2D object embedded within 5 cm of a solid turbid medium (polyurethane) is imaged in transmission with a femtosecond laser and a SPAD camera. b) An example of a time-integrated diffuse image. c) Ground truth image. d) Reconstructed image using the full spatial and temporal information with values under 45\% of maximum set to zero.}
\label{setup_fig}
\end{figure}

In the experimental conditions of Ref.~\cite{Lyons2019b} and considered here, the illumination source is  a pulsed laser with pulse duration 120 fs, centered at 808 nm, 1 W average power at a repetition rate of 80 MHz
and defocused to a beam diameter of 5 cm at the incident surface of the diffuse medium as illustrated in Fig.~(\ref{setup_fig}a). The medium is formed from two slabs of polyurethane foam, each 2.5 cm thick, between which our 2D object is placed. The object is formed of black tape arranged into various letters and shapes. The absorption and reduced scattering coefficients of the medium are measured to be $\mu_{\text{a}} = 0.09$ cm\textsuperscript{-1} and $\mu_{\text{s}}' = 16.5$ cm\textsuperscript{-1} ($\ell^* = 600$ $\mu$m). 

The transmitted photons are imaged with a 32 $\times$ 32 array of SPAD detectors (commercialised by Photon Force). Each pixel is coupled to its own Time Correlated Single Photon Counting circuit to record the arrival times of the incident photons with a bin width of 55 ps \cite{gariepy2015single}. The total length of the acquired temporal point spread function (TPSF) histogram is limited to 230 time bins (approximately 12.5 ns) to avoid overlap with the signal from successive laser pulses.

Figs.~(\ref{setup_fig}b-d) show an example of a time-integrated image at the sample output, the ground truth of the embedded object and the final image reconstruction, respectively. In this case, the reconstruction algorithm uses all data above the noise floor (100 temporal bins), i.e. the full time-of-flight statistics from every pixel, to reconstruct the image \cite{Lyons2019b}. In the following we review the computational inversion process that allows to retrieve the final image and we then investigate the contribution of different temporal sections of the time-of-flight data towards the final reconstruction.

\section{Inverse Retrieval Method}
The image reconstruction method uses a regularised gradient descent algorithm which minimises the least square error between simulated and experimental data. First the data is pre-processed with background subtraction and pixels with a low number of total counts are removed. 
The retrieval is initiated with a guess for the 2-Dimensional object ($x$) hidden between the two 2.5 cm polyurethane slabs (Fig. \ref{setup_fig}), for which we take the time-integrated image of the camera data.  A time-resolved estimate of the fluence after the first slab is found numerically from Eq.~\eqref{Eq:PDE_phi}, convolved in space and time with the Gaussian laser profile. This is then multiplied pixel-wise with the image of the estimated object to find the new intensity incident on the second slab. The fluence at the output of the second slab, $A(x)$, is calculated with same procedure for the first slab (convolution of intensity at second slab with Eq.~\eqref{Eq:PDE_phi}). Finally, we estimate the residual between this 3-Dimensional data cube and the experimental data, $Y$, obtained from the SPAD array.
We then optimise the regularised least squares problem:
\begin{equation}\label{Eq:obj}
    \min_{x}\{O(x)\} = \min_{x}\Big\{||A(x)-Y||_2^2+\lambda_1||x||_{TV}+\lambda_2||x||_1\Big\}.
\end{equation}
To account for the noise in the data and improve the conditionality of the inverse problem, regularisation factors are chosen \textit{a priori} based on known characteristics of the hidden object. The L1-norm promotes sparsity in the image and the norm of the total-variation (TV-norm) uses the sum of the gradient of the image to promote uniformity whilst preserving sharp step-like features. Each regularisation term has a weighting factor ($\lambda_{1,2}$) to control the influence they have on the objective function.
Every iteration calculates the derivative ($O'(x)$) of Eq.~\eqref{Eq:obj} and updates the estimation of the hidden object by:
\begin{equation}\label{Eq:update}
    x^{(k+1)} = x^{(k)} - \alpha O'(x),
\end{equation}
where $k$ is the iteration number and $\alpha$ is the step size.

\section{Image reconstruction with early and late photons}

\begin{figure}[h!]
\centering\includegraphics[width=11cm]{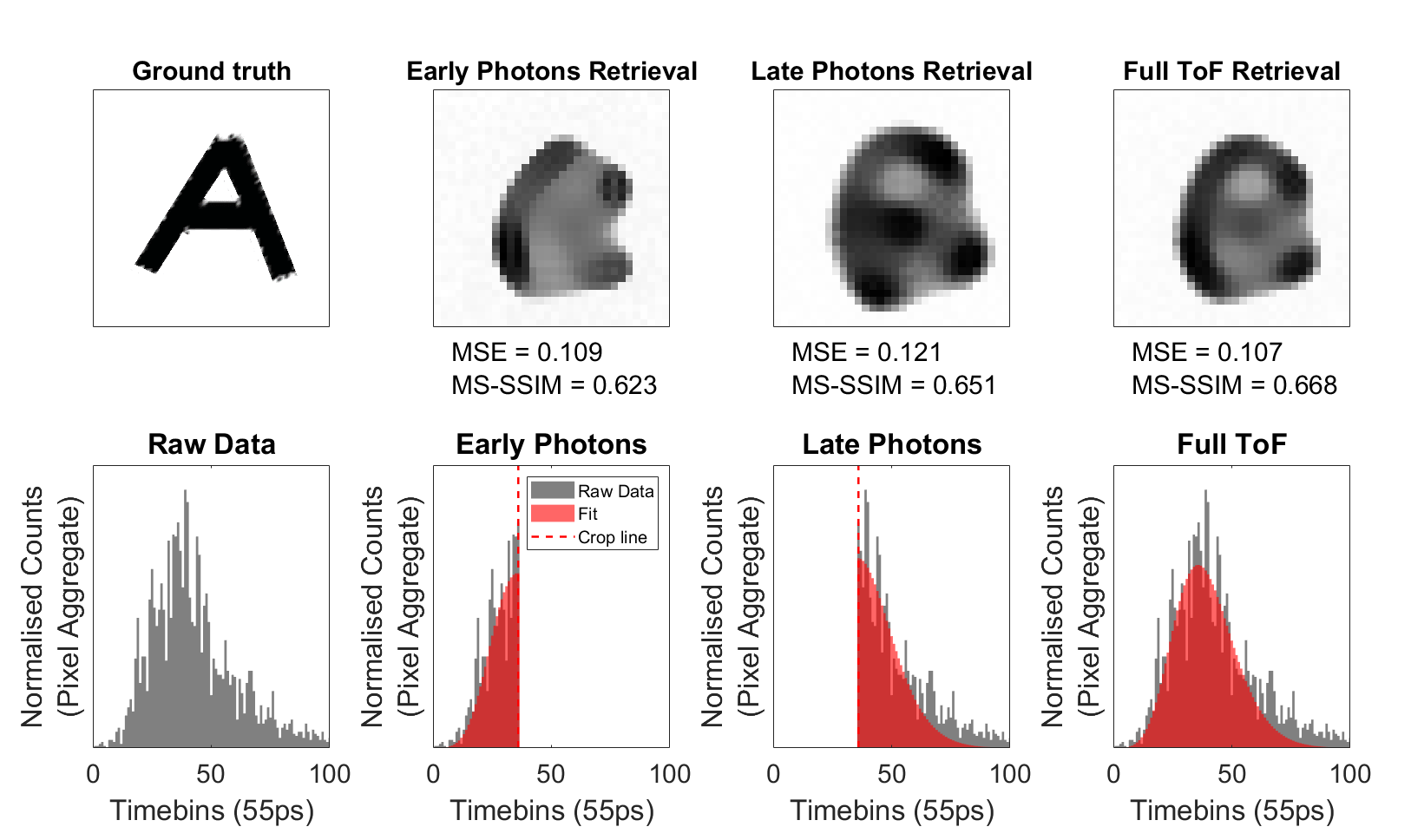}
\caption{Ground truth of hidden object and the optimised reconstructed images (top) for the same experimental data. The data is exclusively cropped for early arriving photons, late arriving photons and the full time of flight (ToF) information (left to right). The mean squared error (MSE) and multi-scale structural similarity (MS-SSIM) metrics are with reference to the ground truth.}
\label{fig:A_data}
\end{figure}

As discussed above, ballistic and snake photons are negligible in our regime. We therefore define "early" and "late" photon arrival times as before and after the peak arrival time of the pixel-aggregated statistics of the temporal histograms. Using the data recorded to reconstruct the image in Fig.~\ref{setup_fig}, we investigate the role of early and late photons by performing image retrievals on specific subsets of the temporal data, for example, selecting only early, only late or including the full set of photon temporal arrival times.

In more detail, we crop the temporal histogram data for each spatial pixel by multiplying with a unit step function, thus selecting only a subset of contiguous time-bins. We then search for the optimised retrieval for the subset of data. Indeed, we found that the quality of the final retrieved image depends on the regularisation parameters and that these parameters need to be optimised when varying the way in which the data is selected.
Therefore, the tunable parameters in the algorithm are the L1-norm regulariser ($\lambda_1$) and the gradient descent step size ($\alpha$). The TV-norm regulariser along with all other parameters have no substantial effect on geometry of the result. The solution space was explored with a parameter grid search (>57 hours computing time) of the L1-norm and step size for every cropped data-set. We then evaluate both the multi-scale structural similarity (MS-SSIM)  \cite{Wang2003} and mean squared error (MSE) of each retrieval with reference to the ground truth: the image with the most qualitative structure to the ground truth was selected and $\alpha$ and $\lambda_2$ were further fine-tuned to optimise the solution. 
The results are shown in Fig.~\ref{fig:A_data}.

We clearly see that using late photons alone (early photons are put to zero) can reproduce an image with similar quality to the ground truth as early photons. Therefore demonstrating that using early-gating does not have a significant advantage in our regime and that late photons have similar information content as early photons. 

\begin{figure}[h!]
    \centering
    \includegraphics[width=0.8\textwidth]{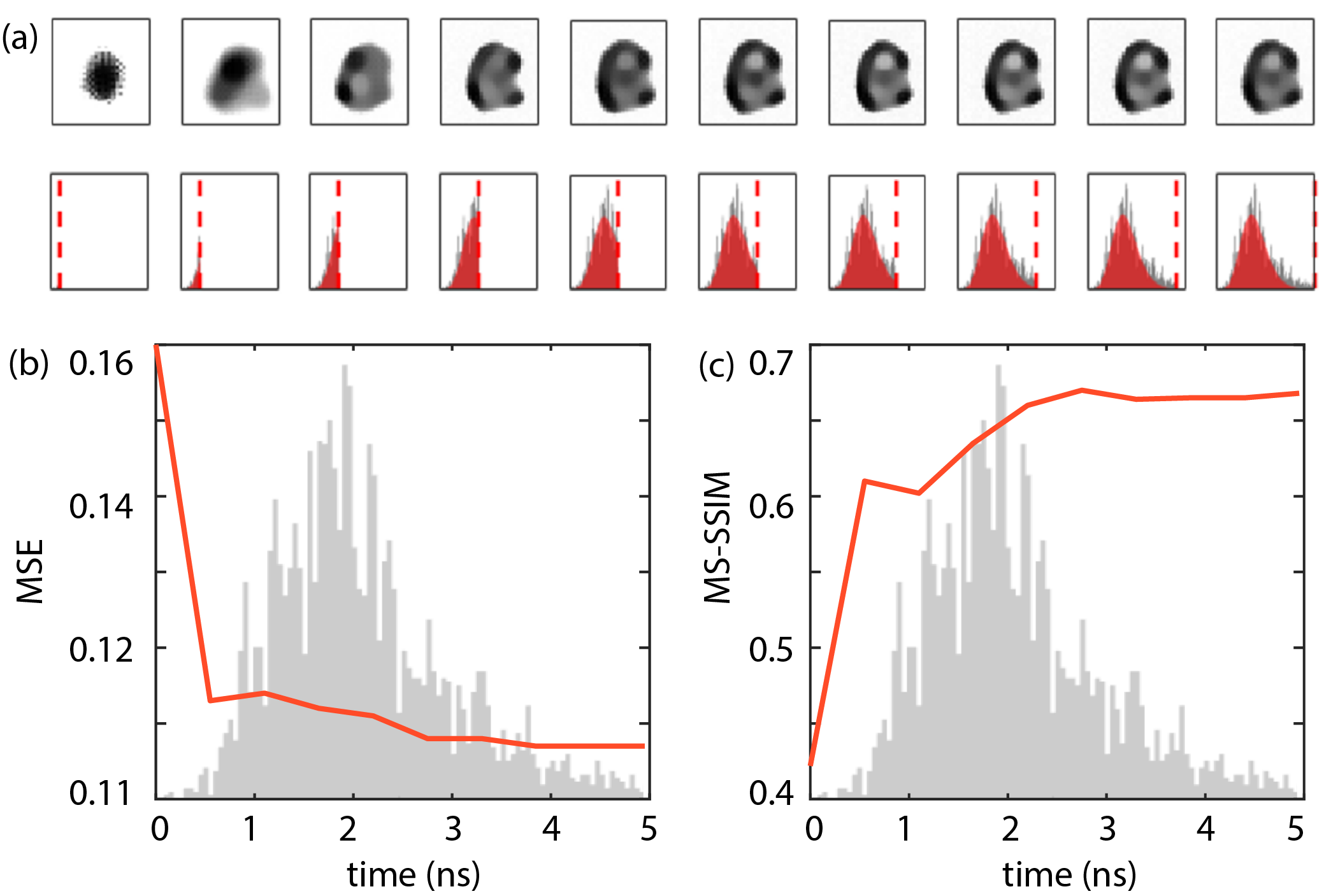}
    \caption{a) The normalised reconstructed image (top row) from the given cropped data (bottom row) when including more gradually late photons (left to right). (b) The mean squared error (MSE) and (c) multi-scale structural similarity (SSIM) metrics calculated from each of the reconstructions shown in (a), with reference to the ground truth. 
    The gray shaded area in all figures shows the actual temporal histogram data.
    The MSE (MS-SSIM) decreases (increases) with increasing late photon inclusion in the data, therefore quantitatively verifying that image retrieval quality is maximised by maximising the time-range of photons included in the computational retrieval.}
    \label{fig:all_scans}
\end{figure}

We investigate this in more detail in  Fig.~\ref{fig:all_scans}. In Fig.~\ref{fig:all_scans}(a) we shift the cropping step-function so as to gradually increase the number of utilised time bins, starting from early time and moving towards late times. 

As before, for each individual retrieval, a full grid search of the regularisation parameters is performed to ensure that in each case, we have the optimal solution. 
Figures~\ref{fig:all_scans}(b) and (c) show the MSE and MS-SSIM for each of these retrievals, plotted versus the time at which the data is cut-off.
These results quantitatively confirm that using  more time bins leads to higher quality image retrievals, confirming that early and late photons play (at least) equally important roles. 

\begin{figure}[h!]
\centering\includegraphics[width=11cm]{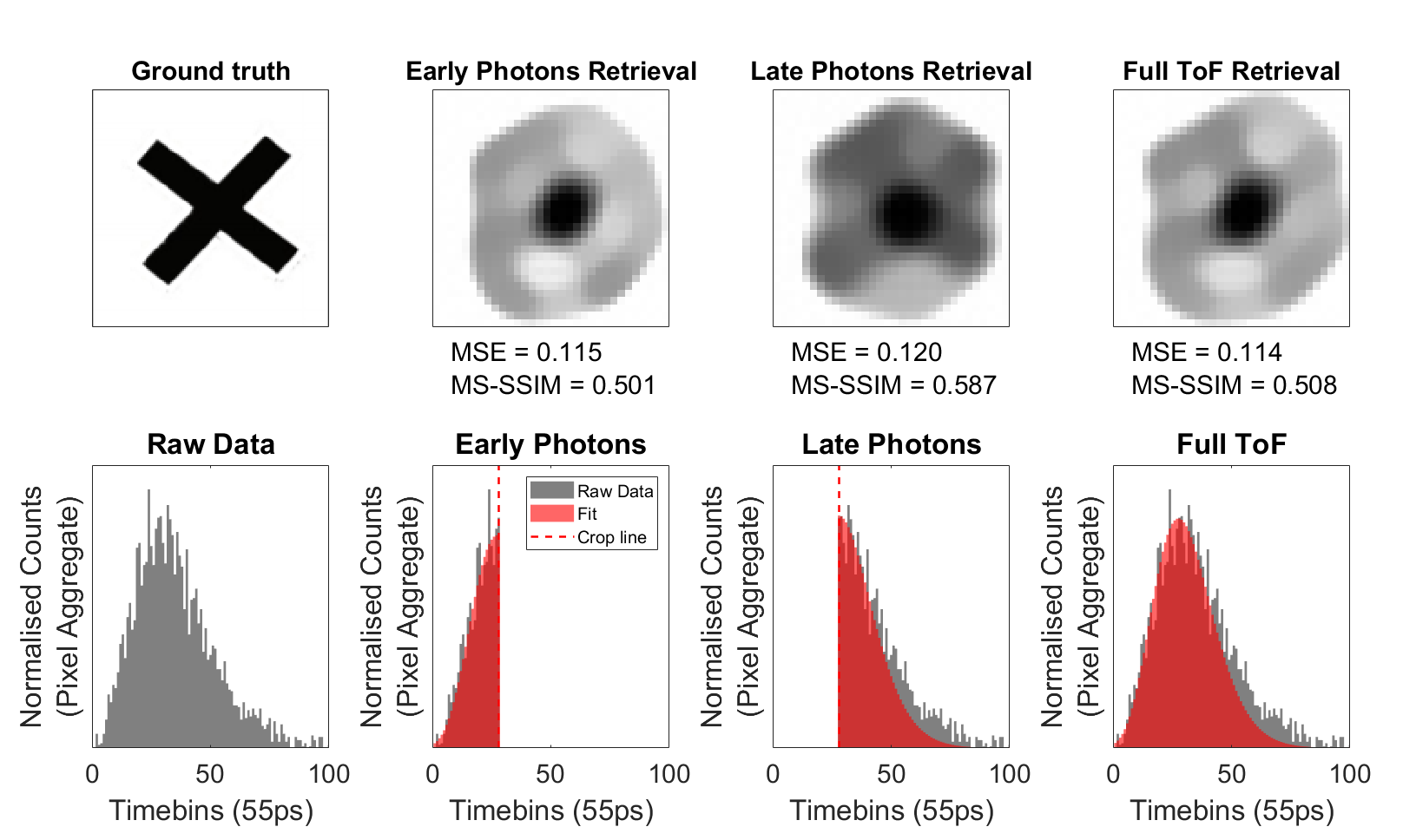}
\caption{Ground truth of hidden object and the optimised reconstructed images (top) for the same experimental data. The data is exclusively cropped for early arriving photons, late arriving photons and the full time of flight (ToF) information (left to right). The mean squared error (MSE) and multi-scale structural similarity (MS-SSIM) metrics are with reference to the ground truth.}
\label{fig:x_data}
\end{figure}

As described above, optimisation of the retrievals and regularisation parameters for each selection of the data is very time consuming. We thus focused here in particular on one specific data set, corresponding to the letter ``A'' object. This was also the more complex shape used in Ref.~\cite{Lyons2019b} and represents the more challenging situation, thus providing insight into the general role of early versus late photons. However, a parameter grid search was also performed with different data for the letter ``X'' in Fig. \ref{fig:x_data}, which also supports the claims in Fig.~\ref{fig:A_data}, that isolating the late photons alone can retrieve an image and all of the reconstructions resemble a similar quality to the ground truth.

The conclusion is that, although in thin scattering media (i.e. $1-10\ell^*$), early photons are scattered less and therefore contain more spatial information, at greater thicknesses (i.e. moving towards $100\ell^*$)  the diversity in the full temporal profile of the photon arrival times provides additional information in the data that is sufficient to constrain the solution space of the increasingly ill-posed inverse problem of imaging through thick, highly scattering media.

\section{Conclusions}

We explore the impact of the arrival time of the detected photons on image reconstruction for time-of-flight diffuse optical imaging. In the regime of imaging through highly scattering media, above 80 transport mean free path lengths, we find that there are neither ballistic nor snake photons recorded by the detector and the gain in spatial information from time-gating early photons is negligible. Therefore, in this regime all of the detected photons are diffuse and contribute comparably to image reconstruction. Late photons, despite containing no direct image information as a consequence of complex propagation paths through the material, do contain information about interactions with absorbing regions in the temporal statistics measured at the detector. Thus the entire temporal profile of the transmitted light should be sampled for high quality image reconstructions. We show that image quality does not degrade when including late photons; rather, reconstructions can be performed even without including any early photons. This indicates that in the highly diffusive regime, with applications for imaging deep inside biological tissues, TCSPC or scanning gate detection methods followed by computational reconstruction that leverages the full temporal information can provide significantly improved image reconstruction. 

\section*{Acknowledgments}
The authors acknowledge funding from  EPSRC (UK, grant no. EP/T00097X/1).  DF is supported by the Royal Academy of Engineering under the Chairs in Emerging Technologies scheme.  JR is supported by the EPSRC CDT in Intelligent Sensing and Measurement, Grant Number EP/L016753/1.
FT is supported by  Amazon  and  EPSRC  grant EP/M01326X/1.

\section*{Disclosures}

The authors declare no conflict of interest.

\bibliographystyle{ieeetr}
\bibliography{diffuse_ref}

\end{document}